\documentclass{article}
\title{Abstraction, Explanation, and Effective Field Theories}
\author{Martin King}
\date{\today}
\usepackage{natbib}
\usepackage{amsmath,amssymb}
\usepackage{tabto}
\usepackage{graphicx}

\begin{document}

\maketitle

 {\Large \center DRAFT COPY, DO NOT CITE, DO NOT READ CAREFULLY}

\begin{abstract}
Effective field theories (EFTs) are widely considered by physicists to be explanatory and to be the appropriate frameworks for modelling various phenomena at different scales.
At the same time, they are known to be approximate, restricted, and merely effective, and thus, examining them can provide a means of getting traction on philosophical issues such as idealisation, abstraction, and the veridicality of representations in explanation. This paper casts EFTs as \textit{abstract} models of a more fundamental theory that retain all and only the relevant aspects for a given explanandum. I describe abstraction as a process that can preserve explanation top-down from an independently explanatory fundamental theory to an effective theory. Thus the paper aims to show how abstract models, like EFTs, can function as explanatory stand-ins for more fundamental models, something often taken to be unproblematic. 
 
\end{abstract}
\pagebreak

\section{Introduction}

Effective field theories (EFTs) are very popular in physics and are increasingly recognized as important to understand in philosophy. 
This has been taken so far as to claim that there are no fundamental theories and that all theories are effective field theories \citep{cao1993}. 
EFTs are widely considered by physicists to be explanatory and to be the appropriate frameworks for modelling various phenomena at different scales.
They are far more tractable than full theories and are pragmatically indispensable. 
At the same time, they are known to be non-fundamental, to only be approximate, to have limited domains of applicability, and in general to be `merely effective'. 
Thus, even if the empirical justification of their use in physics is not lacking, examining them can provide a means of getting traction on philosophical issues about scientific realism \citep{fraser2009,Rivat2020-RIVPFO,wallace2006,williams2018} and inter-theory relations \citep{bain2013,batterman2019,Butterfield_2010,Crowther2016-CROESU}, as well as about idealisation \citep{batterman09,batterrice}, abstraction, and the veridicality of representations.
This latter set of issues is what this paper will focus on, in particular, in the context of scientific explanations. 
What I will show in the paper is that EFTs can be understood as abstract models of more fundamental models
\footnote{As I will be moving back and forth between theories termed models, such as the standard model, and models termed theories, such as effective field theories, I will not distinguish here between models and theories and use the words interchangeably as appropriate by what is customary.} and because of this, insight can be gained into why they should be considered explanatory in certain cases. 
This is not a novel view of EFTs---rather the paper will develop the notion of abstraction and abstract model and make explicit something that is widely assumed, viz. that an EFT can stand in for an explanation from a fundamental theory. 
By looking at the role of the full theory in justifying the idealising assumptions in building and applying the EFT, I present a way to understand what one could call a proxy, or Ersatz, explanation.

In philosophy of science, abstraction is not as often discussed as its sibling, idealisation, but it has much longer history.
Abstraction was an important process for Aristotle as it was the means by which we get knowledge of various concepts and universals \citep{Baeck2014}. 
The intellect removes details from the perceptions of concrete physical objects and retains general properties. 
The intellect can abstract from abstractions, ultimately leading to knowledge of the purely abstract, i.e. of mathematics. 
Abstraction in the context of the philosophy of science is related to this notion, but distinct as it is properly about modelling rather than concept formation.
Abstraction is often characterised as the omission of factual details and contrasted with idealisation, which is the introduction of falsehoods. 
Abstraction is necessarily tied to idealisation and I like to think of it as a second step---one cannot abstract some feature away unless one has already justified an idealisation, i.e. that its removal makes no significant difference. 
One needs to know with respect to what some abstraction makes no difference and so I will characterise abstraction as relative to a given explanandum and an abstract explanation as relative to the explanation of the same explanandum from a more fundamental model.
What I argue is that when there is an explanation from a more fundamental model $M$ and if another model $M'$ is merely an abstraction of that model and can still predict the explanandum, then $M'$ can stand in for $M$ explanatorily.
Importantly, $M$ does a lot of work to justify the idealisations that permit the omissions in $M'$. 
This is not an argument that EFTs provide stand-alone explanations, or that they satisfy particular criteria for explanation, or allow the modelling of novel phenomena, such as argued by \citep{Franklinknox2018}.
I will not be arguing that abstract models provide better explanations, novel explanations, or more optimised explanations \citep{strevens08}, but aim rather to show one way that they are at all explanatory, viz. when they harmlessly stand in for explanations from more fundamental models.

In the following section, I begin with a brief review of the relevant literature on scientific explanation and introduce the problem of veridicality and the notions of abstraction and idealisation that some have used to address it. 
I then describe the process of abstraction as a way of preserving explanation from a fundamental model to a coarse-grained model.
The idea is that an abstract model retains all and only the relevant aspects of a fundamental model for the explanation of a given phenomenon. 
In Section~\ref{sec:top_down}, I put this process to work in explaining the lifetime of the muon from Fermi theory, which I describe as an abstract model of the Standard Model (SM).
I finish the section by outlining some of the ways in which this process would not apply and hence would not preserve explanation, by looking at how this very same process is used to search for new physics.  
In this case, even though the process is formally the same, what one has is a fine-graining rather than a course-graining and here no explanation can be preserved. 

\section{Scientific Explanations and Veridicality} 
\label{sec:explanations}

It is well-established in the philosophy of science literature that scientific theories do not directly contact the world. 
Nancy Cartwright was among the first to really emphasise this and she says ``the route from theory to reality is from theory to model, and then from model to phenomenological law. The phenomenological laws are indeed true of the objects in reality---or might be; but the fundamental laws are true only of objects in the model'' \citeyear[p.~4]{cartwright83}. 
The importance of model mediation is also centre-stage in many more contemporary accounts of models and explanations, such as the \textit{model as mediators} approach outlined by \citet{morganmorrison}, where many models are seen as constructed partly from scientific theories, partly from experimental data, and with external input from modelling practices, such as idealisation, approximation, abstraction, background knowledge, analogy, intuition, and so on. 
This introduces a kind of dilemma for explanation.
We want to think about theories as explanatory, but in practice they are insufficient to capture phenomena directly.
On the other hand, models can describe the phenomena but there are necessarily truth-compromising modelling practices involved.
In other words, how can models explain if they are not veridical?
An explanation of why something happened, or happens, needs to get relevant parts of the story right. 
Something would not be an explanation if it was plausible but factually false. 
In some sense, the falsehoods involved must be harmless.
Many have offered accounts of how models explain, regardless of, or in some cases, because of, the falsehoods involved.

Falsehoods in scientific modelling are typically known as \emph{idealisations}. 
The literature on idealisation is vast and diverse.
Some divide idealisation into three types \citep{weisberg07}, some into six \citep{mcmullin85}.
Some do not distinguish abstraction from idealisation and some contrast idealisation only with approximation, such as \citet{norton2012}.\footnote{\citet{norton2012} refers to an idealisation as a proxy system rather than an inexact description, and while he and others, such as \citet{Butterfield_2010} have shown that limit properties and limit systems may diverge, I will not reserve idealisation for whole systems.}
I can only briefly address some of the contributions to this in the context of explanation where it is most relevant here. 
McMullin was the first to introduce a scheme for idealisation and explanation.   
His stance was that only certain idealisations, called \textit{Galilean} idealisations, could be involved if a model is to be explanatory. 
Galilean idealisations are ones which, when introduced, could be smoothly de-idealised such that one can recover the real system. 
These could include assuming that air resistance on falling body is negligible, that the mass of a pendulum's string is insignificant to its motion, or assuming that the imperfections on an incline plane will not affect the motion of bodies rolled down it, and so on. 
Non-Galilean idealisations are ones for which this process does not work, such as cases where there is a singular limit.\footnote{Explanations and idealisations involving singular limits has been discussed extensively by Batterman \citet[see e.g.][]{batterman02a,batterman05a}.}
At a singular limit, the behaviour of the system is radically different than it is as it approaches the limit, such that one cannot smoothly de-idealise up to and across the limit. 
This precludes recovering the system by de-idealising.

Hempel was explicitly against the possibility that falsehoods could explain, and made the truth of the explanans an explicit requirement for explanation \citep{hempel65}.
The requirement for truth or veridicality, has been softened in much contemporary literature to include approximate truth or to only require truth (or approximate truth) of the aspects of the model that matter for the explanation.   
Others, such as Bokulich have championed the explanatory power of falsehoods, even non-Galilean idealisations \citeyear{boku11,boku12,boku16}.
Bokulich calls these falsehoods \textit{fictions} and argues that, in contrast to various primarily ontic views on explanation, models with fictions can be explanatory.\footnote{Bokulich's notion of fiction seems to apply to elements of models rather than models themselves. The latter view one finds for example in \citep{frigg2009,friggnguyen}, but this is a stance on what models are and is not directly related to our discussion of abstractions and idealisations.}

Abstraction is often thought of as either the inverse of idealisation or as a kind of idealisation itself.
There are many other important processes in modelling and even neatly separating idealisation and abstraction is itself an idealisation.
However typically, idealisation is taken to be the introduction of features or properties that are not present in a target system, while abstraction is the omission or elimination of features that are present in a target system.
We can think of an idealisation as an assumption of a falsehood and abstraction as a removal a truth.
Thus, in contrast to the list above, abstractions would involve the removal of the term for air resistance, setting a pendulum string's mass to zero, the modelling of an incline plane as perfectly flat, and so on.
These cannot truly be separated as a distinct processes from the idealisation, but conceptually it can be useful to distinguish the justification and adoption of idealising assumptions, and the process of their removal from the model. 
The abstract model then, features a selection or subset of those of the full model, as determined by what is relevant for the modelling/explanatory purpose.  
\citet{chakravartty2001} articulates such a view of abstraction.
He describes it as ``a process whereby only some of the potentially many relevant factors or parameters present in reality are built-in to a model concerned with a particular class of phenomena'' (p.~327).
The literature largely agrees that abstraction can be done in at least two broad ways.
\citet{ordorica2016}, as well as \citet{haug2011} and others, have divided abstraction into two distinct process: \textit{abstraction by omission} and \textit{abstraction by aggregation}.\footnote{One notable exception is \citet{janssonsaatsi2019}, who characterise abstraction as a kind of independence from physical structure or law, rather than an omission of detail. For them, explanations from abstract models are not special and are still a function of counterfactual dependence information.}

To see abstraction by omission most clearly imagine that model $M$ is, as far as is scientifically feasible, a complete and accurate description of a real world system, with many details and descriptions included that have no bearing on the derivation of the explanandum phenomenon $E$.
For instance, when explaining why the period of a real, concrete pendulum is 1s, one can (and should) omit descriptions of its colour, its composition, the temperature of the air, and so on, as these can be shown to be completely irrelevant to pendulum's period. 
One can determine this by performing experiments, or simulations, and seeing what changes do and do not make a difference.
But one can also do this by comparing the abstract model with the more fundamental model. 
One can use the more fundamental model to justify the idealisation that some element of the concrete system makes no difference and then abstract it away.
Sometimes, the omission does not matter for a given explanandum, but could matter for others. 
Including the mass of the pendulum's string or the string's elasticity could make a difference to some explananda, but not to others, depending on the system.
This is why abstraction should be thought of as a process conducted relative to a given explanandum.

In addition to this, I think one can subdivide another kind of abstraction by omission. 
One can also omit, not just whole elements or aspects of the model, but some amount of detail or precision by \emph{approximating}. 
This occurs when one rounds off values, or takes only leading terms in an expansion. 
One idealises that the system is exact in some way, e.g. that a given distance is exactly 3m rather than 3.0000001m.
One again uses the explanation from more fundamental model to justify the idealisation and then abstracts away details or levels of precision that make no difference to the explanandum.

Abstraction by aggregation is the treatment of two or more objects, concepts, causes, etc. as one.
It is a kind of coarse-graining where one describes the system using fewer, larger components.
This is the kind of abstraction that one does when one takes information about the averages of particles' momenta in a gas and represents them as pressure, or averages of kinetic energy as temperature, or averages of masses of particles as the center of mass of a body, and so on.
Differences in the states of the fundamental model, or microstates, are not considered. 
Where these differences do not affect the explanandum, one can omit them without losing the explanation. 
In this kind of abstraction, one is effectively changing the scale of the description of the system. 
Changing the scale is not a simple process and likely involves a great deal of idealising assumptions and multiple abstractions. 
In fact, this is what we will see in the case studies below. 
With EFTs, one uses the renormalization group equations redefine a theory at a different scale, but this is a complex procedure. 
We will break it down into a few non-technical steps and highlight the role the full theory plays in justifying the idealising assumptions that allow one to end up with an abstract model.
For some explananda, this scale change will not be problematic and may be an appropriate way to model the system, in the sense that the model at the new scale provides relevant information about possible changes to the explanandum, for example. 
It is the abstract model's independence or autonomy from the microstates that allows the explanation to work at that level.

For clarity, we can now characterise what it means to be an \textbf{abstract model}. 
\begin{quote}
a model $M'$ is an abstraction of another model $M$ if one can get to $M'$ from $M$, using these abstraction processes outlined above
\end{quote}
A model is abstract only relative to another model, which for lack of a better term we can call full or fundamental\footnote{I use the term `fundamental' cautiously here, because it too must be understood as a relative term, contrary, I think, to its common usage.}, which may itself be an abstraction of another model, and so on. 
Models may of course be written down without having been through the process of idealisation and abstraction---model builders do not always, and realistically very rarely, start from a full model or theory and then only make certain kinds of well-justified moves into to arrive at their final model. 
It is really a problem for philosophers to investigate whether the steps taken in a reconstruction can be justified such that explanatory ability is preserved from one model to another. 

Let us make the following claim about explanations from abstract models: 
\begin{itemize}
	\item If a model $M$ explains some phenomenon $E$, then $M'$ also explains $E$ if
	\begin{enumerate}
		\item $M'$ can still derive $E$ and
		\item $M'$ is an abstraction of $M$
	\end{enumerate}
\end{itemize} 
In very general terms, what allows this claim to be made is that the information that is abstracted away is irrelevant to the derivation of the explanandum and we are assured of the irrelevancy by retaining the derivation.\footnote{I do not use the term `derive' in a strictly logical sense, but rather in the sense that it follows from a model together with some initial conditions, as will be made clearer in the case studies below.}
The fundamental model provides essential information as it is the benchmark for what can and cannot be removed safely. 
The idea is to preserve whatever it is, causal or otherwise, that the fundamental explanatory model gets right about the target explanandum. 
Taken together with the description of an abstract model, there emerges an asymmetry or indeterminateness between the abstract and the fundamental. 
One can start from the same model $M$ and arrive at two very different abstract models $M_1'$ and $M_2'$ through the abstraction process if one has a different explanandum.
If the explanation is \textit{that} a phenomenon happens, rather than precisely \textit{how} it happens, or if the explanation has a different contrast class, then different abstractions and different amounts of abstraction can take place while the explanandum can be derived.
If an explanandum is very sensitive to changes in the microstates of the fundamental model, then perhaps almost no abstraction can maintain the explanation.

This has a reasonable conclusion that there may be different models at different levels of abstraction that can nonetheless be the appropriate levels of description for various explanandum phenomena.
The claim also makes clear that this condition is sufficient for explanations of $E$, but not necessary for them. 
There are other reasons why one could consider $M'$ explanatory of $E$, but demonstrating that it is an abstraction of $M$ is sufficient to say that it explains $E$ if $M$ does. 
My claim is that the reason that it can be said to explain is that it is standing in for the full explanatory model in an appropriate way, by preserving what the fundamental model is getting right about the explanandum. 
Clarifying some of the appropriate ways is a central aim of the paper. 

There is a worry that merely performing these abstraction processes is not sufficient to guarantee explanation. 
Let me outline a few ways in which the resulting model would not be explanatory that follow from above claim.
This process will not result in an explanatory model when the underlying or fundamental theory i) is not known, ii) is not capable of deriving the phenomenon, iii) is not explanatory of that phenomenon, or iv) if additional steps/assumptions are required to recover the explanandum from the abstracted model. 
These will be covered in more detail in the final case study below.
They are meant to act as a kind of safety switch.
By over-abstracting and over-generalising, one runs the risk of moving to a level of description where the explanandum may not be derived quantitatively---such as if the numerical prediction cannot be made sufficiently precise for the explanandum, or if the relevant changes in the microstates have been averaged over.
In this case, $M'$ may still explain $E$, but if one or more of i--iv is the case, then another argument for its explanatoriness would have to be presented.

This discussion of abstraction is reminiscent of the eliminative procedure in Strevens' account of explanation. 
For \citet{strevens04,strevens08}, only veridical causes are permissible in explanations, yet he stresses the importance of higher-level and abstract models in explanations.
The basic idea of Strevens' account is to begin with the fundamental description of the system and to keep \textit{all and only} the elements of the model that are required for the explanandum. 
While for Strevens' this picks out the relevant causal difference makers, which may not be appropriate here, I think the basic idea is appropriate for discussing how effective field theories may explain, and in general how abstract models may stand-in for explanations from fundamental models. 

In the remainder of the paper, I will look at effective field theories as abstractions in order to help understand why and under what conditions they can be explanatory.
I will be discussing EFTs in the context of particle physics where they are most common.  
EFTs have been much discussed in the literature and many debates about them have raged. 
EFTs were brought to the attention of philosophers first through \citet{teller89,cao1993,huggetweingard} and \citet{hartmann2001}.
The discussion of EFTs in philosophy is alive and well today \citep[see e.g.][for an excellent survey]{Rivat2020-RIVPFO}, but typically concerns realism and related issues, rather than explanation.
Very briefly, EFTs are coarse-grained theories that are demonstrably accurate (and putatively explanatory) below a certain high-energy limit $\Lambda$, called the ultra-violet (UV) cut off.
I say that an EFT is coarse-grained, because it is restricted to a longer length scale.
The details of the short-range, high-energy physics are negligible for the EFT.
The interactions of heavy particles ($ m > \Lambda $), for example, are `encoded' by direct, contact interactions of light, low-energy particles.
In terms of explanation, the important and distinguishing feature of an EFT is that it includes all and only the relevant degrees of freedom for the explanandum. 

EFTs come in two main types: top-down and bottom-up.\footnote{They are also divided according to whether they are Wilsonian or continuum EFTs and some, such as \citep{bain2013a}, have argued that this makes an important distinction, which may help to resolve some of the EFT debates.}
What these mean will be made clearer with examples in the following section, but for now the following will suffice.
In our context, a top-down EFT takes the SM as a fundamental theory and is constructed to effectively give the same results for some phenomenon at an energy scale lower than that of the fundamental theory.
By contrast, a bottom-up EFT is used to search for new physics, and as such, takes established physics (the SM) as effective theory of some unknown fundamental theory. 
The energy scale of the bottom-up EFT is higher than the SM, and in the top-down, it is lower. 
The idea of using a bottom-up EFT is to identify the effects of new physics and constrain possible models beyond the Standard Model (BSM). 
Which model counts as fundamental and which as effective is relative: the SM can be both depending on whether one is explaining low-energy phenomena or searching for higher energy theories beyond the Standard Model. 

The popularity and importance of EFTs make them an interesting class of models to study. 
Characterising EFTs as abstract models through this process can be useful in understanding why and under what circumstances they can be explanatory. 
The idea of fictions, misrepresentions, and the explanatory power of falsehoods is of particular relevance in discussing effective field theories, because they are known to be only effective theories, but are nonetheless considered to be explanatory by physicists and philosophers in many cases. 
As I have discussed with abstraction, fine- and coarse-grain are relative notions for EFTs and here too, there is an asymmetry.
The information in a more detailed, fine-grained model is sufficient to formulate an abstract and coarse-grained version of that model.
But beginning from an abstract model one cannot determine which detailed model it is an abstraction of. 
There are many UV-completions for a given EFT, hence the difficulty of using this approach to find new physics. 
In this sense, the abstract supervenes on the detailed model and one can move from a more detailed model to coarse-grained one, but not vice versa.
This is why top-down explanation works.

In the following section, I show that a case can be made that in some circumstances an EFT can be considered explanatory, by showing that one can move from an explanatory theory to an EFT with these explanation preserving moves.
The idea is to including first \emph{all} and then \emph{only} the relevant elements of the model. 
However, these circumstances preclude this extending to bottom-up effective field theories. 
This does not imply that bottom-up EFTs cannot be explanatory, but merely that if they are it is not for this reason. 
In fact, given that the SM is widely considered to be explanatory and to be an effective theory itself, despite the fact that we do not know the more fundamental theory, it is very plausible that in certain cases a different argument could be given to demonstrate in virtue of what it is explanatory.

\section{Top-down Explanations} 
\label{sec:top_down}

In this section, I review two EFTs.
I first review a top-down EFT to show that they can be explanatory since they can be thought of as abstractions of an explanatory theory.
I then briefly describe a bottom-up EFT and show that no such story can be told there, and hence if it is to be explanatory, it must be for different reasons. 
My claim is that EFTs can be explanatory if it can be shown that they borrow their explanatory power from some known fundamental theory that is independently explanatory, namely, the Standard Model of particle physics. 
That the SM is itself explanatory is widely accepted and I will simply assume this here. 
This was explicitly argued in \citep{king2020}, where the SM's precise confirmation singles it out as providing the best explanations we have of all observations of elementary particle physics---it is an extremely precise and well-confirmed theory that has a global scope unifying many different kinds of phenomena.
Thus, while the SM may be considered to be an EFT by many, it has many virtuous features that render it independently explanatory (by this I mean that it can be argued to be explanatory without appeal to a more fundamental theory of which it is an abstraction).

\subsection{Case Study: Muon Decay}
A quintessential effective field theory is the Fermi theory, which was developed in the early 1930s as a model for beta decay.
Prior to its development, the continuous energy spectrum from $\beta$ decays indicated that this process violated the principle of the conservation of energy. 
In order to remedy this, Fermi, in 1933, proposed the existence of a light neutral particle, the neutrino ($\nu$).
He described $\beta$ decay as a process whereby a neutron transitions into a proton via the emission of a electron and a neutrino, $n \to p + e +\bar{\nu}_e$.
In this way, both charge and energy could be conserved, and Fermi was able to quantify the lifetime of the neutrino and determine the shape of the $\beta$ ray emission spectrum. 
Today the theory can be used to calculate many more phenomena, such as the decay of muons, particles which were not even known at the time. 
I will take the decay of muons as our explanandum phenomenon $E$ in this case.

Today, Fermi theory is known to be an EFT of the weak interaction of the SM---it is a coarse-grained, low-energy description of the processes that takes place in the SM.
The interactions of the SM electroweak theory are mediated by additional heavy particles, the $W^\pm$ and $Z^0$ bosons.
Nonetheless, for certain processes, like the decay of the muon, Fermi theory can be an adequate description. 
There may be virtues of Fermi theory that render it explanatory on some accounts, but here I would only like to show that it can explain $E$ because i) it is an abstraction of the SM and ii) the SM explains $E$. 
Here, I will simply assume ii), and focus on i).
If we take the process of muon decay to be our explanandum, we will need to show that Fermi theory is an abstract model of the SM (in particular, a coarse-grained model), and that one can still predicts muon decay from Fermi theory. 

To avoid an overly technical description, let us largely follow the general procedure for developing EFTs described in several lecture series \citep{brehmer,georgieft,kaplan,manohar2018}.
This involves a four-step process\footnote{I have divided the first step of the process as described in lectures into two distinct steps for clarity.}:
\begin{enumerate}
	\item Choose an energy scale for the explanandum
	\item Define the field content of the model
	\item Impose symmetries observed
	\item Impose a counting scheme
\end{enumerate}
The first two steps are to set the scale of the abstract model and ensure that one includes all the relevant aspects of the full model that would be required. 
Steps 3 and 4 are to reduce this to all and only the relevant aspects for the explanandum; paring the full model down to an abstract model.
What one is doing here is changing the scale of the model by redefining the parameters to take the cut off into account.
It is a process known as renormalisation and much has been written about this in philosophy.
There are many great sources that treat this in some detail, so I will focus on the digestible presentation developed in the lectures just mentioned.   
Very briefly, the renormalization group equations tell you how the theory's parameters change as the energy is varied.
Counter terms must be introduced to absorb the effects of the high energy physics, but they can be harmlessly removed if they are irrelevant to the explanandum.
EFTs make a great case study for abstraction since the process features abstraction by omission (removal of irrelevant terms), by aggregation (in the parameter redefinitions), and approximation (in truncating the expansion).

We should begin with some preliminaries.
The basic object describing a quantum field theory is the action

\begin{equation}
	S=\int{d^4x \mathcal{L}(x)},
\end{equation}
an integral over the Lagrangian density (henceforth Lagrangian), where the Lagrangian is a sum of operators $\mathcal{O_i}$ with coefficients $\alpha_i$ called couplings

\begin{equation}
	\mathcal{L} = \sum^i \alpha_i \mathcal{O}_i
\end{equation}
The operators are combinations of fields and derivatives of the fields evaluated at a point $x$. 
The couplings can be split into a dimensionless constant called the Wilson coefficient $c_i$, and some powers of the mass scale $\Lambda$, $\alpha_i = \dfrac{c_i}{\Lambda}$.
The terms of the Lagrangian are either kinematic terms and mass terms, which describe the the theory's fields, or interaction terms that describe how those fields interact.
The EFT we are aiming at is a Lagrangian valid at low energy, that includes only light degrees of freedom, and which can accurately derive our explanandum of muon decay. 
What we want is a sum of the new interaction terms, seeing as the kinetic and mass terms are described by the SM and can be bracketed here ($\mathcal{L}_{SM}$).

\begin{equation} \label{efteq}
	\mathcal{L}_{EFT}= \mathcal{L}_{SM} + \sum_{i=1}\dfrac{c^d_i}{\Lambda^{d-4}}O^d_i
\end{equation}
Each term here is indexed by its mass dimension $d$, which is a key property of the operator determined by summing the mass dimensions of the fields involved in the interaction.
In natural units of $\hbar = c = 1 $, mass = energy = length$^{-1}$ and everything dimensionful can be given a dimension of mass. 
Each term in a Lagrangian in 4-dimensional spacetime must have mass dimension 4. 
As the dimensionality of the operator increases, so must the power of the cut off in order that the term remains dimension-4, as in Eq.~\ref{efteq} . 
This suppresses the effect of higher dimension terms in the Lagrangian and allows one to reasonably truncate the perturbative expansion.
Let us now construct the EFT from the full model. 
One can follow the full physics derivation in the sources above. 

\noindent
\textbf{1. Specify the energy scale.}
The first step in getting this Lagrangian is to specify a relevant energy scale below which the EFT will operate, determined by the relevant scale for the explanandum. 
The cut off should be above the scale of the muon ($\sim$106~MeV), since it is a light particle we wish the EFT to describe explicitly. 
We should take the scale to be below the mass of the weak bosons, $W^\pm$ and $Z^0$ (83~GeV and 91~GeV), otherwise we will end up with the full SM description.
This is critical, because the separation of scales is what allows us to idealise that the details of the high-energy physics are irrelevant to the explanandum.

\noindent
\textbf{2. Define the field content.}
As a second step, we integrate out all the fields whose masses are higher than the cut off. 
One must define all the relevant fields, namely those with $m \ll \Lambda$, and include all the possible terms describing interactions between these fields at all orders.
In our case, the particle content of the EFT will then be the leptons and quarks of the SM (without the top).
This specifies content of the theory, but it also involves an infinite number of interaction terms, including many that violate the symmetries of the SM and those of arbitrarily high mass dimension.
Now that all the relevant degrees of freedom are included, one needs to pare down such that one is left with all \emph{and only} the relevant degrees of freedom.

\noindent
\textbf{3. Impose symmetries.}
Depending on the explanandum many different symmetries could be imposed, such as gauge symmetries, spacetime symmetries, flavour symmetries, etc. that we know will be observed by these interactions at this low energy.
We know such things because we have precision measurements of predictions from the fundamental model.
In our EFT of the weak interaction, we will impose Lorentz invariance, the conservation of electric charge, colour charge, lepton and baryon number. 
We then abstract away by omission the terms that violate these symmetries. 
This still leaves an infinite number of terms, because there are operators at all orders. 
In order to prevent this, we must truncate the expansion of operators.

\begin{figure}[t]	\label{muon}
\includegraphics[width=0.9\textwidth]{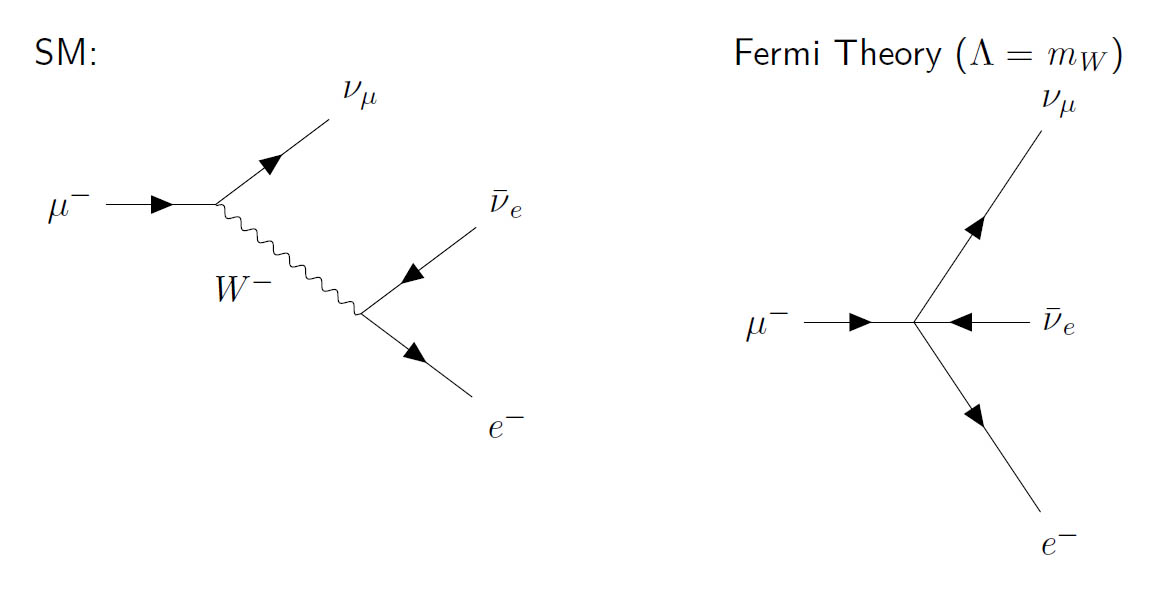}
\caption{Muon decay process as described by the Standard Model (left) and the Fermi theory (right).}
\end{figure}

\noindent
\textbf{4. Impose a counting scheme.}
As the mass dimension of the operators increase, their effects are suppressed by increasing powers of the cut off, and so they quickly become irrelevant. 
Here, we have an idealising assumption that further precision is not necessary.
One can impose a maximum operator dimension and we will truncate after the lowest order, effectively approximating. 
Fermions have mass dimension 3/2, so operators with three fermion fields violate lepton conservation and Lorentz invariance, so four is the smallest number.
Four fermions gives a dimension-6 operator for muon decay:

\begin{equation} \label{muoneq}
	\mathcal{L} = -\frac{4G_F}{\sqrt{2}}(\bar{e} \gamma^\mu P_L \nu_e)(\bar{\nu}_\mu \gamma^\mu P_L \mu)
\end{equation}
This describes an interaction between these four fermions that encodes the physics processes of the SM. 
Diagram \ref{muon} shows how the point-like interaction between four fermions stands in for the vector boson mediations of the SM. 
From the effective Lagrangian in Eq.~\ref{muoneq} one can predict that the muon decays and calculate its lifetime. 
At lowest order it is given by the following

\begin{equation}
	\Gamma_\mu = \dfrac{G^2_F m^5_\mu}{192 \pi^3}
\end{equation}
Without using the full calculation from the SM, we have been able to predict the explanandum with good accuracy.
The EFT error can be shown to be on the order of the ratio of the energy scales $\frac{\mu}{W}$ or about 0.0011\%. 
This error also shows why the separation of scales is so important for modelling as an EFT as it provides a measure of the approximation. 
Because one can perform the full calculation from the SM, one can verify that the EFT prediction is close enough for the explanandum.
In our abstract model, there is no mention of the W and Z bosons, the interaction term is dim-6 and non-renormalisable. 
However, the SM allows us to justify the idealising assumptions in the construction of the EFT. 



 
The artificially reconstructed process is as such: one begins with an explanandum phenomenon and some fundamental or full description of a system that can derive it---this guarantees that you have \emph{all} the relevant features for the explanandum.
Then, one pares down irrelevant aspects to an abstracted model featuring \emph{all and only} the relevant features of the system given the explanandum. 
In both cases, we can think of these steps as explanation preserving, in that they can be understood as merely and harmlessly abstracting from the full explanatory model. 
If they were not harmless, the derivation would no longer work.
Whatever the fundamental model gets right in its explanation is encoded into the abstract model and nothing new is introduced. 
However, this argument is not a blank cheque for claiming that all EFTs are explanatory, and certainly not that EFTs are independently explanatory.
There are a few ways in which this could fail to preserve explanation and I will turn to some of these in the following section.

\subsection{Bottom-up Exploration} 
\label{sub:bottom_up}


The steps described in the previous section preclude explanation from a large group of EFTs that are widely used in searches for new physics.
These are called \emph{bottom-up} EFTs in contrast to the \emph{top-down} EFTs discussed above.
One of the ways in which these EFTs are distinct involves their role in explanation, which I will demonstrate by briefly discussing the standard model effective field theory (SMEFT). 
Let us once again construct our EFT and I will review some of the ways in which the attempt at explanation from an abstract model may fail.

The idea behind using a bottom-up effective field theory is to consider the known theory (SM) as an effective version of some higher energy (UV) theory we have yet to discover. 
Deviations from the known theory, or the lack thereof, can put constraints on the possible UV theories.
There are two principle assumptions in the SM-EFT approach, viz. that the scale of new physics is sufficiently separated from the scale of known physics, and second, that the symmetries of the SM that are still approximately observed. 
These assumptions are well warranted. 
In fact, as the LHC data continues to indicate no deviations from the SM, the further back the scale of new physics is pushed and the more justified that assumption becomes.
New physics is likely sufficiently decoupled from the SM. 
In short, the bottom-up EFT procedure is to expand the SM with additional higher-order terms that are not in the SM, but which would encode the effects of new physics on SM-observables. 
One then performs global fits to see if there are any significant non-zero coefficients on some of the operators, which would indicate new physics.
Let us construct this bottom-up EFT using the now familiar 4 steps. 

\noindent
\textbf{1. Specify the energy scale.} The first step is to define the relevant energy scale. 
The SMEFT is used to find new physics at scales higher than the SM, $\Lambda_{EFT} < \Lambda_{SM}$, say at the order of a few TeV. 

\noindent
\textbf{2. Define the field content.}
The energy level of the cut off provides us with our field content, which in this case is all of the particles of the SM.
With this list of fields, we need to first consider every possible interaction term between them at all orders. 

\noindent
\textbf{3. Impose symmetries.}
Again, most of the terms up to this point are impossible or irrelevant, since they either have minuscule effects or violate symmetries that are observed in SM physics. 
We can impose the SM symmetries which we expect will likely be approximately respected by the new physics.
Typically, this is the SM gauge symmetries, SU(3)$\times$SU(2)$\times$U(1), lepton and baryon number (though these latter two are accidental symmetries of the SM). 
At this point, we have what is often referred to as the Standard Model Effective Field Theory:

\begin{equation}
	\mathcal{L}_{SMEFT} =  SM + \sum \frac{c^6}{\Lambda^2}\mathcal{O}^6	+ \sum \frac{c^8}{\Lambda^4}\mathcal{O}^8 + \ldots
\end{equation} 
This is too many terms, once again, to be tractable for physicists, so they typically focus on those of lowest dimension. 

\noindent
\textbf{4. Impose a counting scheme.}
Lastly, we can once again truncate the expansion and focus on terms of the next highest order (strongest effect), which is dimension-6.\footnote{While there are terms of dimension-5, like the Weinberg operator, these violate lepton number conservation and so are rarely studied.}
At this point we are left with a finite number of terms: the SM terms, and the sum of terms with dimension-6 operators.
We can refer to this object as the dim-6 SMEFT. 
Unlike the previous cases, however it is not a manageable number. 
In fact, there are 2499 operators with mass dimension 6 that satisfy all these constraints, each with a coupling constant that is a free parameter. 
This should set off alarm bells as a model with that many free parameters can be fit to anything. 

In a recent paper, we have explored the SMEFT in more detail and argued that there are a few reasons that it is quite distinct from full BSM models and simplified models. 
Part of these reasons are also relevant to why the SMEFT should not be considered explanatory by the abstraction story and I will briefly review these here.
Given the condition on explanation above, there are two things that can make it non-explanatory: 1) there is no explanandum; 2) if one is articulated, there could be two problems: either i) extra steps were taken to arrive there from bottom-up; or ii) the fundamental theory was not explanatory to begin with so there is no explanation to preserve.
As for 1), it should be noted that we have performed these four step without the precise articulation of an explanandum.
All we used was the vague notion that there would be new physics at a scale higher than that of the SM (and a bit higher than has already been excluded at the LHC).
Therefore, it should be pretty clear that the SMEFT is not explanatory, because after these steps one cannot derive $E$, because there is no $E$.
This might seem trivial, since one could just articulate some BSM effect, call that our explanandum $E$ and accommodate it with the SMEFT.
This brings with it two additional issues.

(i) Firstly, if one wants to derive this BSM explanandum $E$, then one needs to take a series of extra steps after our step 4 to continue to pare down the Lagrangian from its current 2499 terms to all and only the relevant ones for the explanandum. 
For example, one could focus on only a particular sector; one could consider only operators that have already been found to have some non-zero coefficient (as global fits have already been performed); or one could select operators that should have the largest effects on a given process of interest. 
Through these extra down-selection steps, one could arrive at an effective Lagrangian with a manageable number of terms that could accommodate some possible deviation. 
The problem with this is that these steps will be arbitrary and additional idealising assumptions will have to be included---assumptions that are not guaranteed or even permitted by the fundamental model (the effect is beyond the SM). 
The requirement that the loss of detail is irrelevant can no longer be guaranteed. 
(ii) Secondly, even if the explanandum is some new BSM effect, then the derived EFT is still not explanatory, given what I have argued here, because the SM cannot explain that effect. 
If it is some BSM effect, then there is no SM explanation and nothing to preserve as one moves to the abstract model. 


The culprit in this case should be pretty clear: when we specified a higher energy scale for the EFT, we began asking for a fine-grained picture. 
We switched from removing irrelevant detail to asking what the fine-grained detail could look like, given what we know about the coarse-grained theory. 
Unlike on CSI, one cannot simply enhance a grainy photo without inventing detail.
The argument here is about the preservation of explanatory information, which is not symmetrical between fundamental and abstract.

Whether there is an explanation to preserve depends on what one takes as the fundamental model, but what is fundamental and effective is relative. 
So could one not start from a higher energy theory than the SM, for example, Supersymmetry?
Yes, certainly.
One would arrive at what would be a top-down EFT, even though it could be considered bottom-up from the SM.
It could be that what was once the result of a bottom-up process is later a model that can be given a top-down justification. 
What is top-down and what is bottom-up is also relative. 
If one thinks SUSY explains $E$, then some top-down EFT abstracted from SUSY that can derive $E$ would also explain $E$, if one accepts the argument of this paper. 
However, in order for the top-down EFT to explain the BSM effect, one would have to consider the full UV theory, supersymmetry or whatever it may be, to be explanatory. 
At the moment, there is little grounds for thinking that supersymmetry provides actual explanations. 
Supersymmetry provides at best potential, or candidate, explanations, as argued by \citep{king2020}, since it has yet to be experimentally confirmed.

There is a kind of continuity and an arbitrariness between top-down and bottom-up EFTs. 
One can think of top-down and bottom-up EFTs as different stages of the same process, but these stages that make a big epistemic difference.
In this section, I have been arguing that not all EFTs are explanatory because some involve fine-graining, rather than coarse-graining, which is not a move that preserves explanatory information. 
Whether one should consider the EFT explanatory has to do with explanatory judgements about the fundamental theory. 
Again, I have not argued that bottom-up EFTs cannot be explanatory, but that if they are, then it will not simply be because they are coarse-grained versions of known explanations.

\section{Conclusion} 
\label{sec:conclusion}

I have here focused on cases in particle physics since that is where effective field theories are most at home, but this argument for explanation by abstraction would apply in other cases as well. 
One could provide a similar story about explanations of gravitational phenomena from a Newtonian model by showing how it is an abstraction of GR; about explanations of why the sky is blue using Raleigh scattering as an EFT of the SM; or about how the Ising model provides an explanation of ferromagnetism without having to derive that phenomenon directly from a more fundamental theory. 
This argument is most at home in physics due to the mathematical nature of the process and the reliance on the explanatory features of a global theory. 
In other disciplines, practicing scientists may be less concerned with the role of theory and the theoretical justifications of their modelling assumptions, but this story of explanation by abstraction may still apply. 
There is no need to see this as an attempt to ground all explanations in fundamental physics, but merely to show that a model can be explanatory by being appropriately related to a more fundamental theory that is empirically well confirmed and unifies a greater range of different kinds of phenomena.

This paper has not attempted to argue that it is only under these circumstances that EFTs are explanatory, rather that if the abstraction process can be accomplished, then the model can justifiably be said to be explanatory---it is a sufficiency claim, rather than a necessity claim. 
Many models that would not count as explanatory by abstraction may indeed be explanatory for other reasons. 
Our best theories of fundamental physics are considered to be EFTs by many physicists and there are good grounds for counting them as explanatory, at least in the sense that they are the best explanations we have. 
Again, while there is a certain continuity between top-down and bottom-up EFTs, the difference I wish to highlight has more to do with epistemology than physics.
This paper began by investigating some of the problems that veridicality poses for explanation. 
Models that are merely effective, or phenomenological, are not veridical, but are often still considered explanatory. 
What I have provided here is a story about why and when they should be considered explanatory. 
I have shown that one can understand EFTs as providing a kind of proxy explanation if one sees them as models abstracted from more fundamental models that are independently explanatory. 
It is often assumed that where there is an explanation from a fundamental theory, that an EFT also explains. 
I have attempted to defend this by showing how the EFT stands in for the fundamental model and to highlight the role the fundamental model plays in justifying the idealising assumptions of the EFT.


\bibliography{../../library}
\bibliographystyle{apalike}

\end{document}